\documentclass[prl,twocolumn,floatfix,showpacs ]{revtex4-1}
\UseRawInputEncoding
\usepackage{amsmath}
\usepackage{amssymb}
\usepackage{graphicx}
\usepackage{dcolumn}
\usepackage{bm}
\usepackage[colorlinks=true,linkcolor=blue,anchorcolor=blue, citecolor=cyan,urlcolor=cyan]{hyperref}
\usepackage[mathlines]{lineno}
\usepackage{ulem}
\usepackage{epstopdf}
\usepackage{tabularx} 
\begin{document}
\title{Hidden  half-metallicity}
\author{San-Dong Guo}
\email{sandongyuwang@163.com}
\affiliation{School of Electronic Engineering, Xi'an University of Posts and Telecommunications, Xi'an 710121, China}
\author{Pan Zhou}
\email{zhoupan71234@xtu.edu.cn}
\affiliation{Key Laboratory of Low Dimensional Materials and Application Technology of Ministry of Education, School of Materials Science and Engineering, Xiangtan University, Xiangtan 411105, China}
\begin{abstract}
Half-metals, featuring ideal 100\% spin polarization, are widely regarded as key materials for spintronic and quantum technologies; however, the half-metallic state is intrinsically fragile, as it relies on a delicate balance of exchange splitting and band filling and is therefore highly susceptible to disorder, external perturbations, and thermal effects. Here we introduce the concept of hidden half-metallicity, whereby the global electronic structure of a symmetry-enforced net-zero-magnetization magnet is non-half-metallic, while each of its two symmetry-related sectors is individually half-metallic, enabling robust 100\% spin polarization through a layer degree of freedom. Crucially, the vanishing net magnetization of the entire system suppresses stray fields and magnetic instabilities, rendering the half-metallic functionality inherently more robust than in conventional ferromagnetic half-metals. Using first-principles calculations, we demonstrate this mechanism in a $PT$-symmetric bilayer $\mathrm{CrS_2}$, and further show that an external electric field drives the system into a seemingly forbidden fully compensated ferrimagnetic metal in which hidden half-metallicity persists. Finally, we briefly confirm the realization of hidden half-metallicity in altermagnets, establishing a general paradigm for stabilizing half-metallic behavior by embedding it in symmetry-protected hidden sectors and opening a new route toward the design and discovery of unprecedented half-metallic phases.
\end{abstract}
\maketitle
\textcolor[rgb]{0.00,0.00,1.00}{\textbf{Introduction.---}}
Half-metals, in which one spin channel is metallic while the other is insulating, exhibit ideal 100\% spin polarization and are therefore regarded as key materials for high-performance spintronic devices~\cite{q0}. Such perfect spin selectivity can give rise to ultrahigh tunneling magnetoresistance (TMR) and enable substantial improvements in magnetic storage density and energy efficiency~\cite{q1}. Despite these appealing properties, the half-metallic state is intrinsically fragile, as it relies on a delicate balance of exchange splitting and band filling and is thus highly susceptible to disorder, structural imperfections, external perturbations, and thermal fluctuations. As a result, many materials predicted to be half-metallic under ideal conditions fail to retain full spin polarization in realistic experimental environments. Collinear magnets can be broadly classified into net-zero-magnetization and nonzero-magnetization systems~\cite{k4,k5,f4,aplgsd}. While half-metallicity can be realized in ferromagnets, ferrimagnets, and fully-compensated ferrimagnets-among which the latter are particularly attractive due to their vanishing net magnetic moment and reduced stray fields~\cite{f4,z1,s2,s3},  the symmetry-enforced degeneracy between spin-up and spin-down electronic states rigorously forbids conventional half-metallicity in $PT$-antiferromagnets ($PT$) of space inversion symmetry ($P$) and time-reversal symmetry ($T$)) and altermagnets.

\indent The concept of hidden physics offers a promising strategy to overcome such symmetry-imposed constraints. A paradigmatic example is hidden spin polarization in globally centrosymmetric systems composed of locally non-centrosymmetric sectors, where spin-split electronic states exist in real space but cancel in the bulk~\cite{h3,h4,h5,h7,h12}. This insight has stimulated extensive exploration of a broad class of hidden phenomena~\cite{h12}, including hidden orbital polarization, hidden Berry curvature, and hidden valley polarization~\cite{h121,h122,h123}. More recently, this framework has been extended to magnetic systems~\cite{nc}, leading to the proposal of hidden altermagnetism, in which a globally $PT$-symmetric phase exhibits zero net spin polarization while each symmetry-related sector hosts finite local spin splitting~\cite{f6,f666}. Very recently, hidden altermagnetism has been experimentally confirmed in $\mathrm{Cs_{1-\delta}V_2Te_2O}$ using neutron diffraction and spin-resolved angle-resolved photoemission spectroscopy~\cite{f667}. In parallel, the concept of hidden fully-compensated ferrimagnetism has also been introduced, further enriching the landscape of hidden magnetic states~\cite{f668}.

\indent Motivated by these developments, we explore whether half-metallicity can be realized in a hidden manner in net-zero-magnetization magnets that are globally forbidden by symmetry from hosting conventional half-metallic states. In this work, we propose the concept of hidden half-metallicity, in which two symmetry-related sectors are each individually half-metallic, while their contributions cancel in the global electronic structure, resulting in zero net spin polarization. This unique combination of local 100\% spin polarization and vanishing total magnetization offers an intrinsically more stable alternative to conventional ferromagnetic half-metals. Using first-principles calculations, we demonstrate the feasibility of this concept in a $PT$-symmetric bilayer $\mathrm{CrS_2}$ as a prototypical example.\\
\begin{figure*}[t]
    \centering
    \includegraphics[width=0.80\textwidth]{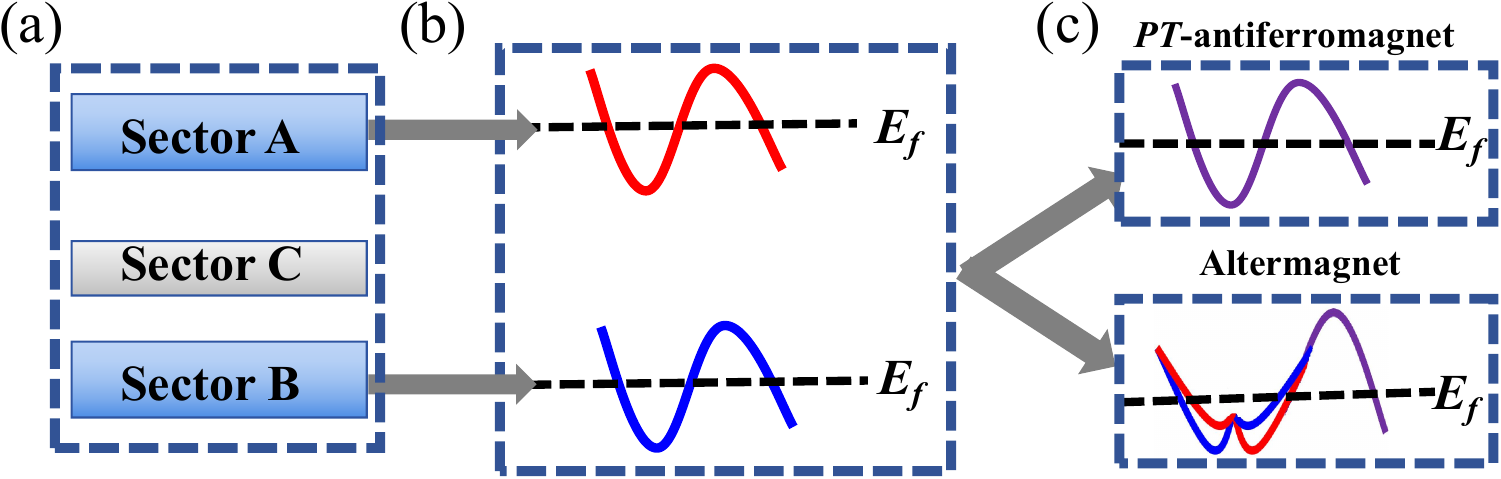}
    \caption{(Color online)(a):this system belongs to the class of symmetry-enforced net-zero-magnetization magnets, encompassing primarily $PT$-antiferromagnet and altermagnet. The system is composed of three sectors: A, B, and C. The electronic states near the Fermi level are primarily contributed by A and B, while those from C lie far from the Fermi energy. (b): the sectors A and B are half-metallic, exhibiting opposite spin polarizations.  (c): viewed globally, the energy band  is either spin-degenerate or exhibits altermagnetic spin-splitting. In (b, c), the blue, red, and purple curves denote the spin-up, spin-down, and spin-degenerate bands, respectively, while the horizontal black dashed line indicates the Fermi level.}\label{a}
\end{figure*}

\textcolor[rgb]{0.00,0.00,1.00}{\textbf{Fundamental mechanism.---}}
For a collinear symmetry-enforced  net-zero-magnetization magnet (for example, $PT$-antiferromagnets and altermagnets), the spin-up and spin-down bands satisfy the relation:
\begin{equation}\label{d-1}
E_{\uparrow}(\vec{k})=[C_2\parallel O]E_{\uparrow}(\vec{k})=E_{\downarrow}(O\vec{k})
\end{equation}
where the  $O$ denotes $P$ or  rotational ($C$) or mirror  ($M$) symmetry, and  the $C_2$ is the two-fold rotation perpendicular to the spin axis in spin space.  According to \autoref{d-1}, we obtain:
\begin{equation}\label{d-2}
g_{\uparrow}(E)=\sum_{\vec{k}}\delta[E-E_{\uparrow}(\vec{k})]=\sum_{\vec{k}}\delta[E-E_{\downarrow}(O\vec{k})]
\end{equation}
As  $\vec{k}$ runs over the entire first Brillouin zone (BZ), $O\vec{k}$ also spans the entire BZ. Therefore, \autoref{d-2} can be rewritten as:
\begin{equation}\label{d-3}
\sum_{\vec{k}}\delta[E-E_{\downarrow}(O\vec{k})]=\sum_{\vec{k}^{\prime}}\delta[E-E_{\downarrow}(\vec{k}^{\prime})]=g_{\downarrow}(E)
\end{equation}
where the $g_{\uparrow}(E)$ and  $g_{\downarrow}(E)$  are defined as spin-up and spin-down densities of states (DOS).  According to \autoref{d-2} and \autoref{d-3}, the $g_{\uparrow}(E)$ and  $g_{\downarrow}(E)$  are identical ($g_{\uparrow}(E)$=$g_{\downarrow}(E)$) in collinear symmetry-enforced net-zero-magnetization magnets.
This implies that half-metallic states are absent in collinear symmetry-enforced net-zero-magnetization magnets, because such states require an imbalance between the spin-up and spin-down  DOS.

Although a collinear  symmetry-enforced net-zero-magnetization magnet cannot be half-metallic as a whole, the local half-metallicity can still emerge once the spatial 'layer'  degree of freedom is introduced.
In this study, we introduce the concept of  hidden  half-metallicity (See \autoref{a}), which posits that the 'local' half-metallicity with only one spin channel crossing the Fermi level can arise, when the specific atomic layer marked with sector A or B  is  ferromagnetic (FM), ferrimagnetic or fully-compensated ferrimagnetic half-metal. In other words, the magnetic atoms in sectors A and B  can be coupled either ferromagnetically  or antiferromagnetically (ferrimagnetic or fully-compensated ferrimagnetic half-metal).
When the magnetic atoms of sectors A and B  are connected by [$C_2$$\parallel$$P$] symmetry, their local spin polarizations are opposite, leading to spin degeneracy. In a two-dimensional (2D) system, the [$C_2$$\parallel$$M_z$] symmetry  can also enforce spin degeneracy  (For  2D systems, the wave vector $k$ only has in-plane
components, and then $E_{\uparrow}(k)$=[$C_2$$\parallel$$M_z$]$E_{\uparrow}(k)$=$E_{\downarrow}(k)$).
If the magnetic atoms in  sectors A and B are linked by a  [$C_2$$\parallel$$C/M$]-symmetric connection, it will give rise to altermagnetic spin-splitting.

Identifying bulk materials that harbor hidden half-metallicity remains a  challenge.
Here, we realize hidden half-metallicity via an alternate bilayer-stacking-engineering strategy.
Employing an analogous construction scheme in ref.\cite{h123}, we can engineer a  bilayer system with [$C_2$$\parallel$$P$]  or  [$C_2$$\parallel$$M_z$]  symmetry by using  the half-metallic monolayer as the basic building unit. For altermagnetic bilayer systems, some general construction strategies have also already been proposed\cite{k8,k9,k10,k11}.
Here, we elucidate hidden half-metallicity in detail by taking the $PT$-symmetric $\mathrm{CrS_2}$ bilayer with  antiferromagnetic (AFM) interlayer coupling  as a representative case.

\begin{figure*}[t]
    \centering
    \includegraphics[width=0.90\textwidth]{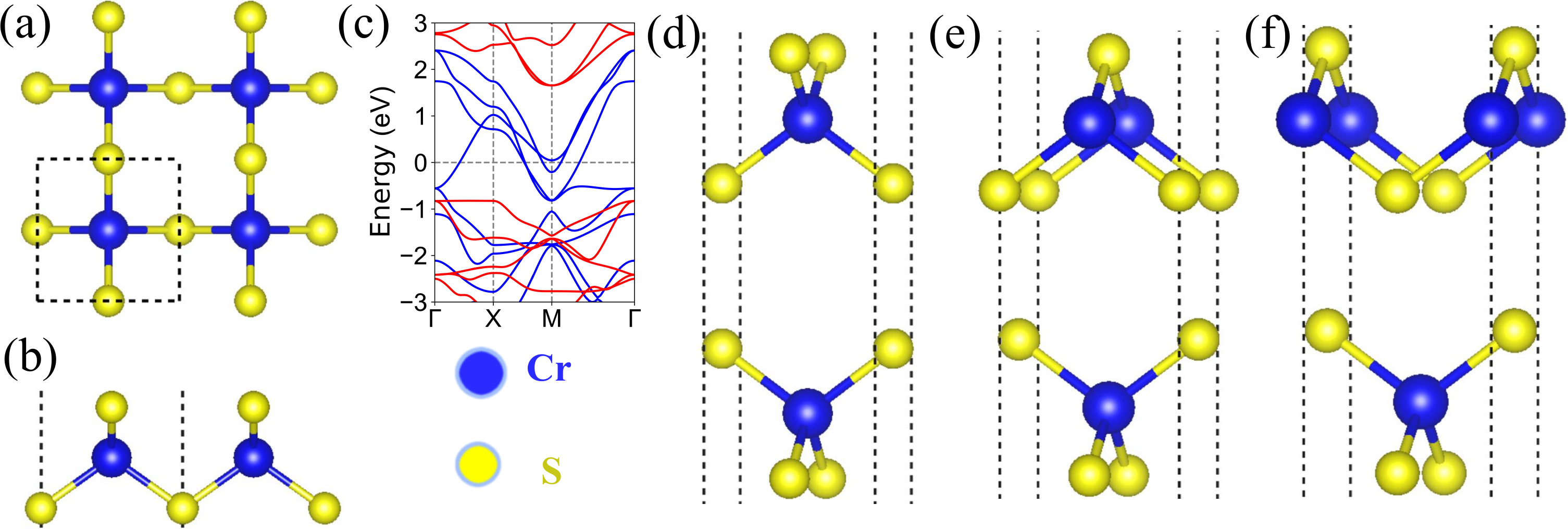}
    \caption{(Color online) For $\mathrm{CrS_2}$,   (a, b , c): the crystal structures and  energy band structures  of  monolayer; (d, e, f): the  AA-, AB-, and AC-stacked bilayer crystal structures.  In  (c), the spin-up
and spin-down channels are depicted in blue and red.  }\label{b}
\end{figure*}

\textcolor[rgb]{0.00,0.00,1.00}{\textbf{Computational detail.---}}
Density functional theory\cite{1,111} calculations are performed by  using the Vienna ab initio simulation package (VASP)\cite{pv1,pv2,pv3} within  the projector augmented-wave (PAW) method. We use the generalized gradient approximation (GGA) of
Perdew, Burke, and Ernzerhof (PBE)\cite{pbe} as the exchange-correlation functional.
A kinetic energy cutoff of 500 eV,  total energy  convergence criterion of  $10^{-8}$ eV and  force convergence criterion of 0.001 $\mathrm{eV\cdot{\AA}^{-1}}$ are adopted.
We add Hubbard correction with $U$=3.00  eV\cite{k12} and  3.68 eV\cite{k13} for $d$-orbitals of Cr and V atoms within the
rotationally invariant approach proposed by Dudarev et al\cite{du}.
The vacuum slab of more than 20 $\mathrm{{\AA}}$ is added to avoid the physical interactions of periodic cells.
 We use a 15$\times$15$\times$1 Monkhorst-Pack $k$-point meshes  to sample the BZ for structure relaxation and electronic structure calculations. The dispersion-corrected DFT-D3 method\cite{dft3} is adopted to describe the van der Waals
interactions.

\begin{table}
    \centering
    \caption{For AA-, AB-, and AC-stacked bilayer $\mathrm{CrS_2}$, the energies (meV) of AFM and FM interlayer couplings  relative to that of AC-stacking with AFM interlayer coupling. }
    \label{tab}
    \begin{tabularx}{0.45\textwidth}{@{\extracolsep{\fill}}cccc} 
        \hline\hline
                 & AA&AB&AC \\ \hline
        AFM & 102.8&54.4&0\\ \hline
       FM &105.6&52.2 &-2.1\\ \hline\hline
    \end{tabularx}
\end{table}

\textcolor[rgb]{0.00,0.00,1.00}{\textbf{Material realization.---}}
Monolayer $\mathrm{CrS_2}$ has been predicted to be stable in  dynamics, thermodynamics, and mechanics\cite{k12}. The $\mathrm{CrS_2}$    contains three atomic sublayers with  one Cr atom sandwiched between two S atomic layers, as shown in \autoref{b} (a, b).  Each Cr atom is connected
with four S atoms, constituting a distorted tetrahedron.
The   $\mathrm{CrS_2}$  crystallizes in the $P\bar{4}m2$ space group (No.115),  lacking  lattice symmetry $P$. To determine the magnetic ground state, we construct three magnetic configurations, including
FM, AFM1 and AFM2 orderings  (see FIG.S1\cite{bc}). Our calculations demonstrate that the FM configuration is the ground state, exhibiting an energy per unit cell that is 414 meV and 242 meV lower than those of  the AFM1 and AFM2 configurations, respectively.  The optimized  lattice constants are $a$=$b$=3.637 $\mathrm{{\AA}}$ with FM ordering.

\autoref{b} (c) presents the spin-resolved energy band structures of monolayer  $\mathrm{CrS_2}$. The spin-up channel intersects the Fermi level, whereas the spin-down channel exhibits an  indirect band gap of 2.48 eV. Consequently, monolayer $\mathrm{CrS_2}$ behaves as a metal for spin-up channel and as a semiconductor for spin-down channel, manifesting intrinsic half-metallicity. The observed half-metallicity is attributed to Hund's-rule splitting of the partially filled $d$-orbitals of the Cr atom, accompanied by weak $d$-$p$ orbital hybridization\cite{k14}. Consequently, charge transport is governed exclusively by the spin-up channel, yielding 100\% spin polarization.

\begin{figure*}[t]
    \centering
    \includegraphics[width=0.90\textwidth]{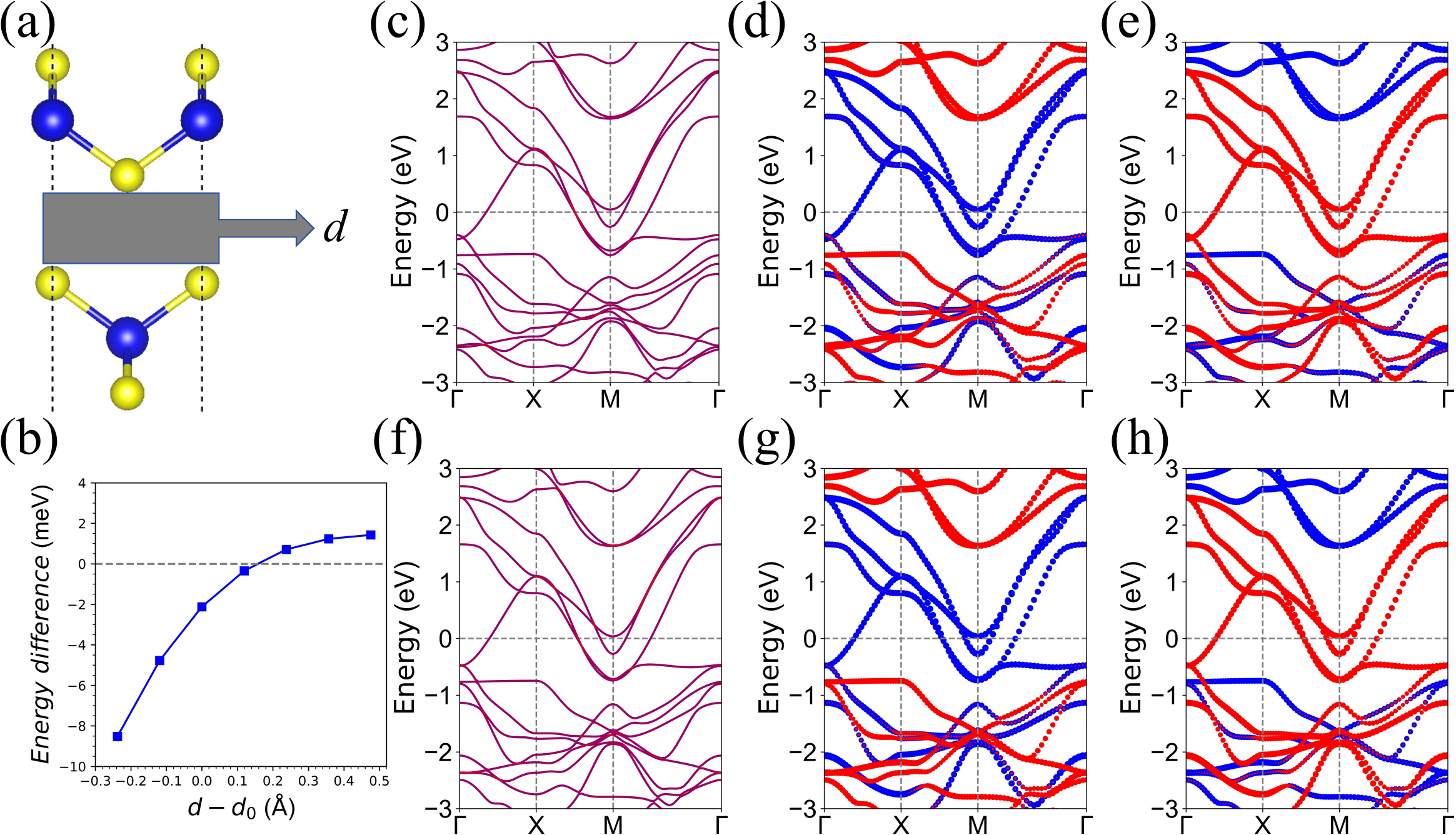}
     \caption{(Color online) For AC-stacked bilayer $\mathrm{CrS_2}$,  (a): the schematic illustration of tuning the interlayer spacing between the lower and upper layers; (b): the energy difference between FM and AFM interlayer couplings versus $d-d_0$, where the $d_0$ denotes the equilibrium interlayer distance;  (c, d, e): the total energy  band structure  along with the spin-resolved projections onto the lower and upper layers  with  the equilibrium interlayer distance;  (f, g, h): the total energy  band structure  along with the spin-resolved projections onto the lower and upper layers  with  $d-d_0$ being  0.48 $\mathrm{{\AA}}$.  In ( c, d, e, f, g, h), the blue, red, and purple curves denote the spin-up, spin-down, and spin-degenerate bands, respectively.  In (d, e,  g, h), the weighting coefficient is proportional to the circle size.}\label{c}
\end{figure*}
\begin{figure}[t]
    \centering
    \includegraphics[width=0.45\textwidth]{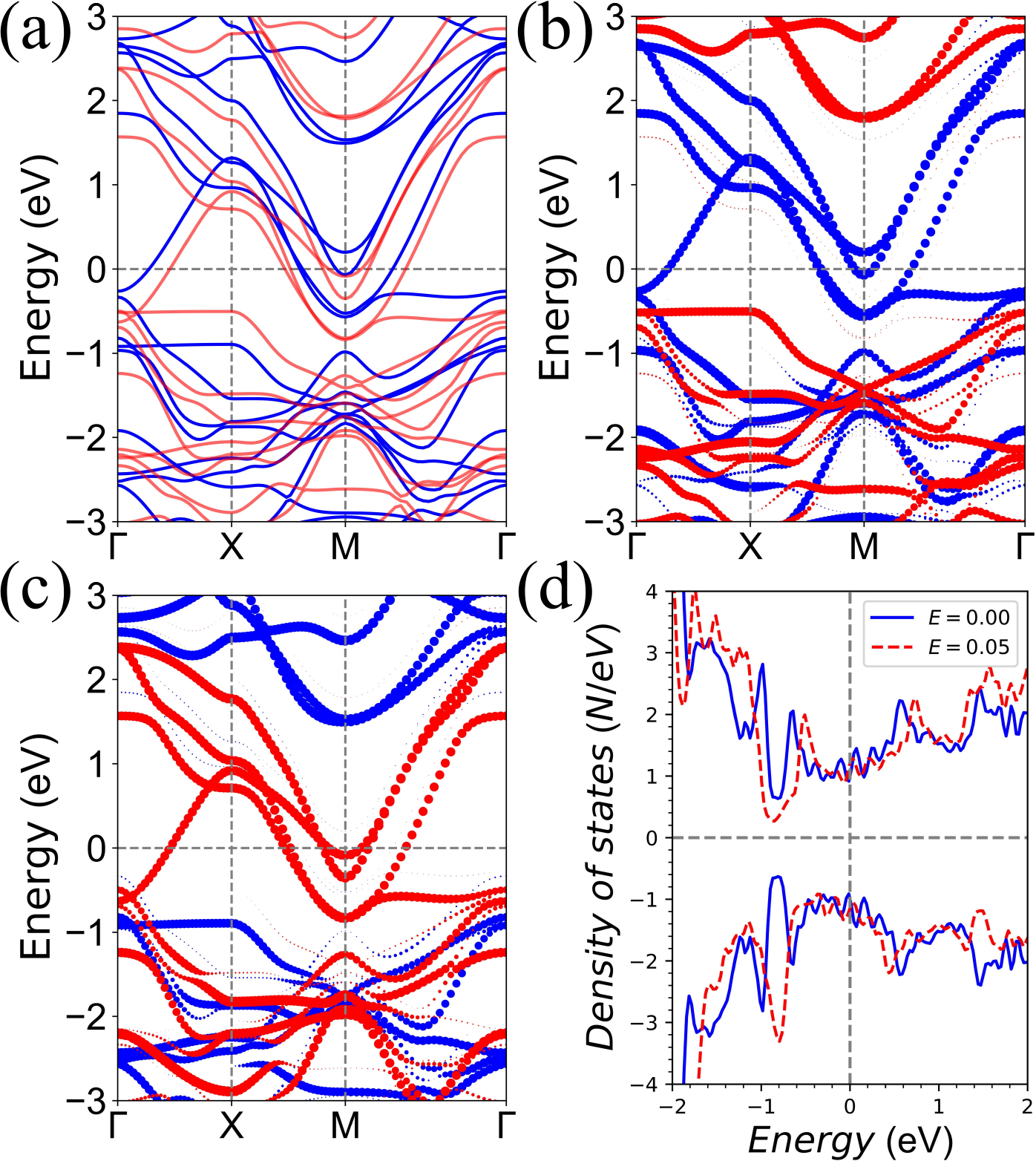}
     \caption{(Color online)  For AC-stacked bilayer $\mathrm{CrS_2}$  with  $d-d_0$ being  0.48 $\mathrm{{\AA}}$, (a, b, c): the total energy  band structure along with the spin-resolved projections onto the lower and upper layers   at $E$=0.05  $\mathrm{V/{\AA}}$;  (d): the  total DOS   at $E$= 0.00 and 0.05  $\mathrm{V/{\AA}}$.  In (a, b, c), the blue, red, and purple curves denote the spin-up, spin-down, and spin-degenerate bands, respectively.  In (b, c), the weighting coefficient is proportional to the circle size.}\label{d}
\end{figure}

Next, we build  $PT$-symmetric bilayer $\mathrm{CrS_2}$ using the construction scheme\cite{h123}. Initially, we take the $\mathrm{CrS_2}$ as the basic building unit, defined as sector B. Through a mirror operation $M_z$, the sector A can be derived from the sector B.
The resulting bilayer (see \autoref{b} (d) as AA-stacking)  possesses not only $M_z$ symmetry but also acquires $P$ symmetry automatically, because monolayer $\mathrm{CrS_2}$ intrinsically exhibits $C_{2z}$ symmetry ($P$=$C_{2z}$$M_z$).
In general, AA-stacking is not the energetically most favorable configuration\cite{f6}.
Based on  AA-stacking, the AB- and AC-stackings  (see \autoref{b} (e) and (f)) can be constructed by translating the upper  sublayer by $\vec{a}$/2 along the $a$-axes and  ($\vec{a}$+$\vec{b}$)/2 along the diagonal direction, respectively.
The AA-, AB- and AC-stackings crystallize in the $Pmmm$ (No.47),  $Pmma$ (No.51) and $Pmmn$ (No.59), and they all have lattice symmetry $P$.

For monolayer $\mathrm{CrS_2}$,  the energies of  AFM1 and AFM2 orderings are several hundred meV above that of FM ordering, whereas the interlayer magnetic interaction is typically smaller than 10 meV\cite{f6,h123,k11}.  We therefore fix the intralayer FM ordering and vary only the interlayer magnetic coupling (FM or AFM) to determine the magnetic ground state of the bilayer. The energies of the three stacking arrangements with the two interlayer magnetic couplings are summarized in \autoref{tab}.
It is found that the AC stacking is significantly lower in energy than both AA and AB stackings, and the following discussion focuses exclusively on the AC-stacked bilayer. Unfortunately, the ground state of AC-stacked bilayer exhibits FM interlayer coupling. Nevertheless, a transition between FM and AFM interlayer coupling can be realized by modulating the interlayer spacing $d$ (see \autoref{c} (a)).  The interlayer spacing can be tuned with  laser shocking (LS) engineering.  In  $\mathrm{FePSe_3/Fe_3GeTe_2}$ heterostructures,  the LS can permanently reduce the interlayer spacing from 7.4  $\mathrm{{\AA}}$ to 4.6  $\mathrm{{\AA}}$\cite{k15}.
The energy difference between FM and AFM interlayer couplings as a function of  $d-d_0$ is plotted in \autoref{c} (b), where the $d_0$ denotes the equilibrium interlayer distance. It is found that reducing the interlayer separation favors FM coupling, whereas expanding the interlayer distance promotes AFM interlayer coupling. When $d-d_0$ exceeds 0.15 $\mathrm{{\AA}}$, the AC-stacked bilayer switches from FM to AFM interlayer coupling.

For comparison, the band structures of the interlayer AFM-coupled AC bilayer are presented in \autoref{c} for both  $d-d_0$= 0 and 0.48 $\mathrm{{\AA}}$, along with the spin-resolved projections onto the lower and upper layer $\mathrm{CrS_2}$.
Owing to $PT$ symmetry, the bands are manifestly spin-degenerate and exhibit no half-metallic character. Nevertheless, when the layer degree of freedom is resolved, the projected band structures of the lower and upper layers clearly reveal that each individual layer retains half-metallic character.
The lower and upper layers exhibit opposite spin polarizations: for the lower-layer, spin-up characteristic band crosses the Fermi level, whereas for the upper-layer, spin-down characteristic band intersects the Fermi level. Consequently, AC-stacked bilayer $\mathrm{CrS_2}$ indeed realizes a hidden half-metallic state.
Given that the ground state of AA-stacked bilayer $\mathrm{CrS_2}$ is intrinsically antiferromagnetically coupled, a hidden half-metallic state can also be achieved by sliding from the AC to the AA stacking configuration (see FIG.S2\cite{bc}).
In inversion-asymmetric crystals, the HSP cannot be directly detected unless $PT$ symmetry is broken. Nonetheless, the ARPES and  spin-resolved ARPES have already experimentally verified the HSP effect in a variety of materials\cite{h4,h5,h7}. Hence, the hidden half-metallicity proposed here is also accessible to experimental confirmation by using  analogous techniques.

A hidden half-metal can be clearly revealed by applying an out-of-plane electric field to break $P$ symmetry. We apply an electric field of $E$=0.05 $\mathrm{V/{\AA}}$ to the AC-stacked bilayer $\mathrm{CrS_2}$   with  $d-d_0$ being  0.48 $\mathrm{{\AA}}$. The calculation results show that it still retains an interlayer AFM ground state, which lies 2.5 meV lower in energy than the interlayer-FM state. The total energy  band structure  along with the spin-resolved projections onto the lower and upper layers are plotted in \autoref{d} (a, b, c). It is clearly seen that the spin degeneracy is lifted due to the broken $PT$ symmetry.  When the direction of the electric field is reversed, the order of the spin-splitting is also reversed. Another noteworthy point is that the total magnetic moment is strictly 0.00 $\mu_B$, indicating that the AC-stacked bilayer $\mathrm{CrS_2}$  under an applied electric field is a fully-compensated ferrimagnetic metal\cite{gsd}. According to the projected band structure, the spin-up and spin-down bands near the Fermi level originate predominantly from the  lower and upper layers, respectively.  By applying an electric field, we can achieve a hidden half-metal in a fully-compensated ferrimagnetic metal. Applying an electric field makes the spin-up and spin-down DOS at the Fermi level differ (see \autoref{d} (d)), so the spin-polarized carriers in the lower and upper layers become distinct. This opens a route to electrically manipulate the spin degree of freedom.

\begin{figure}[t]
    \centering
    \includegraphics[width=0.45\textwidth]{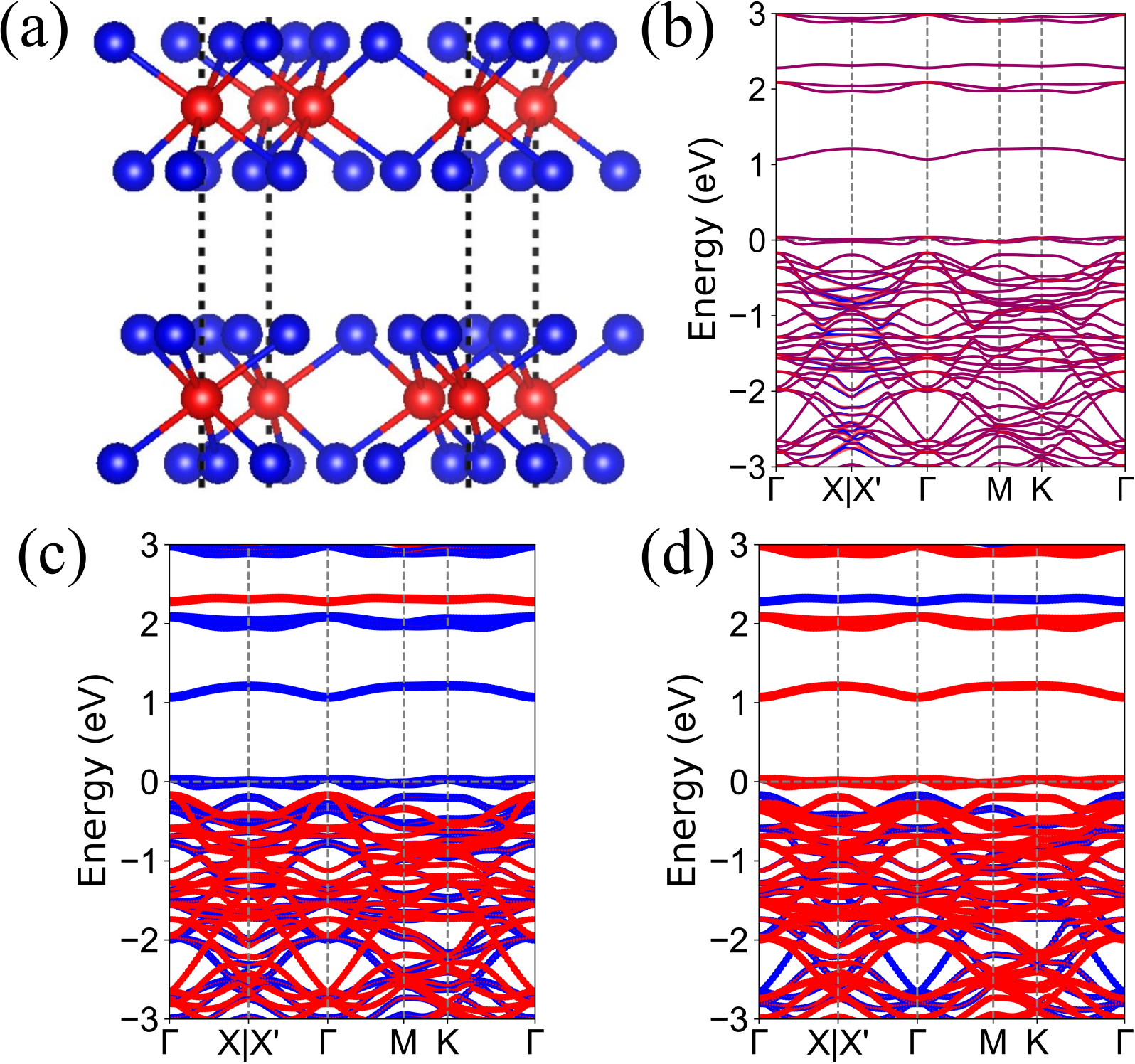}
     \caption{(Color online)  For AB'-stacked bilayer $\mathrm{VI_3}$,  (a): the  crystal structures; (b, c, d): the total energy  band structure  along with the spin-resolved projections onto the lower and upper layers.  In (a), the red and blue balls represent V and I atoms, respectively. In (b, c, d), the blue, red, and purple curves denote the spin-up, spin-down, and spin-degenerate bands, respectively.  In (c, d), the weighting coefficient is proportional to the circle size.}\label{e}
\end{figure}

\textcolor[rgb]{0.00,0.00,1.00}{\textbf{Discussion and Conclusion.---}}
The hidden half-metallicity can also be realized in bilayer altermagnets. Here, we illustrate our idea with an unrealistic AB'-stacked bilayer $\mathrm{VI_3}$ as an example (the term "unrealistic" refers to the fact that its ground state is interlayer FM ordering rather than AFM ordering.).
Monolayer $\mathrm{VI_3}$ has been shown to be a stable FM half-metal\cite{k13}, and its crystal structures  and energy  band structures are presented in FIG.S3\cite{bc}.
Firstly, the AA'-stacked bilayer   is constructed by placing one layer $\mathrm{VI_3}$ on top of
another layer $\mathrm{VI_3}$. The  AB'-stacked bilayer  is
constructed based on the AA' stacking  by sliding the top layer horizontally relative to the bottom
layer with  2/3$\vec{a}$+1/3$\vec{b}$, which is plotted in \autoref{e} (a).
For AB'-stacked bilayer $\mathrm{VI_3}$,   the total energy  band structure along with the spin-resolved projections onto the lower and upper layers are shown in \autoref{e} (b, c, d). To clearly visualize the altermagnetic spin-splitting, an enlarged view of the overall band structure together with the first BZ are plotted in FIG.S4\cite{bc}.
For AB' stacking, the symmetry operations mainly include  $C_{3z}$,  $C_{2x}$,  $C_{2y}$ and  $C_{2xy}$. The $C_{3z}$ symmetry connects atoms within the same spin, while the $C_{2x}$,  $C_{2y}$ and  $C_{2xy}$ connect atoms
with opposite spins,  giving rise to altermagnetic spin-splitting. Although the global band structure shows no half-metallicity, the layer-resolved projected bands unambiguously demonstrate that each individual layer remains half-metallic. Consequently, a hidden half-metal in altermagnets can, in principle, exist.

 In summary, we report the concept of a prospective hidden half-metal that exhibits local 100\% spin polarization.
The system requires spin-antiparallel magnetic atoms to be symmetrically coupled, and  is comprised of individual sectors with  local  half-metallicity.
Taking the $PT$-symmetric bilayer $\mathrm{CrS_2}$  as
a representative, we demonstrate that the hidden half-metallicity can be achieved, and  an  out-of-plane external electric field can be used to tune  layer-resolved spin-polarized carriers.
Our findings open a new avenue for half-metallic research, providing both theoretical guidance and material recipes that will  enable exploration of its emergent physics.

\begin{acknowledgments}
This work is supported by Natural Science Basis Research Plan in Shaanxi Province of China   (2025JC-YBMS-008). We are grateful to Shanxi Supercomputing Center of China, and the calculations were performed on TianHe-2. 
\end{acknowledgments}

\end{document}